\newif\ifuseprd
\newif\iftoomuchdetail
\newif\ifeprint

\useprdtrue 
\toomuchdetailfalse
\eprinttrue

\newif\ifrr
\rrfalse
\newcommand\RR{{\ifrr{RR}\else{Ramond-Ramond (RR)}\fi\global\rrtrue}}

\newif\ifns
\nsfalse
\newcommand\NSNS{{\ifns{NS-NS}\else{Neveu-Schwarz (NS-NS)}\fi\global\nstrue}}

\ifuseprd 
\documentclass[\ifeprint preprint,\fi
tightenlines,floats,prd,eqsecnum,nobibnotes%
,nofootinbib,aps]{revtex4}
\else
\documentclass[final,12pt,letterpaper]{JHEP}
\fi 
\usepackage{amsmath,amssymb}

\allowdisplaybreaks[3]
\newcommand\skipthis[1]{{}}

\makeatletter
\def\@strike{\relax\leavevmode
  \ifmmode
    \expandafter\mathpalette\expandafter\math@strike
  \else
    \expandafter\make@strike
  \fi}
\def\math@strike#1#2{%
  \setbox\z@\hbox{$\m@th#1{#2}$}\fin@strike}
\def\make@strike#1{%
  \setbox\z@\hbox{\color@begingroup#1\color@endgroup}\fin@strike}
\def\fin@strike{%
  \@tempdima\dp\z@
  \@tempdimb\ht\z@
  \lower\@tempdima\hbox{\strike@start}%
  \box\z@
  \raise\@tempdimb\hbox{\strike@end}}
\def\strike@start{\special{ps: %
    currentpoint /starty exch def /startx exch def}}
\def\strike@end{
}
\newcommand\fs{\protect\@strike}

\let\oldAE\AE
\renewcommand\AE{{\ifmmode{\text{\it\oldAE}}\else{\oldAE}\fi}}

\newlength{\parboxlen}
\newcommand\myparbox[3][]{{\settowidth{\parboxlen}{{#2}}%
    \parbox[{#1}]{\parboxlen}{{#3}}}}

\newlength{\notlen}

\newcommand\mathone{{\rlap{\kern .25em l}1}}
\newcommand\one{{\ifmmode{\text{\mathone}}\else{\mathone}\fi}}
\newcommand\proj{{\ensuremath{{{\mathbb P}}}}}
\newcommand\ct[1]{{\ifeprint\ifuseprd{\em{#1}},\else{\sf {#1}},\fi\fi}}
\newcommand\bt[1]{{\em {#1}},}

\chardef\til=`~

\newcommand\half{{\ensuremath{\frac{1}{2}}}}

\newcommand\p{\ensuremath{\partial}}

\newcommand\norm[1]{\ensuremath{\left\|{#1}\right\|}}

\newcommand\transpose{{\ensuremath{\text{\sf T}}}}
\newcommand\field[1]{{\ensuremath{\mathbb{{#1}}}}}
\newcommand\ZZ{{\field{Z}}}

\newcommand\anti[2]{\ensuremath{\left\{{#1},{#2}\right\}}}
\newcommand\com[2]{\ensuremath{\left[{#1},{#2}\right]}}

\newcommand\lie[2]{\ensuremath{\pounds_{{#1}} {#2}}}

\newcommand\apr{{\ensuremath{{\alpha'}}}}

\newcommand\rrf{{\ensuremath{{^{(5)}F}}}}

\providecommand\TABLE[2][tbh]{\begin{table}[#1]\begin{center}{#2}\end{center}
                       \end{table}}
\providecommand\squeezetable{\small} 
\providecommand\toprule{\hline} 
\providecommand\botrule{\hline} 
\providecommand\colrule{\hline} 
\providecommand\putabstract[1]{\ifuseprd\begin{abstract} {#1} \end{abstract}%
                           \else \abstract{{#1}} \fi}
\providecommand\plb[3]{{Phys.\ Lett.\ B {\bf {#1}}, {#3} ({#2})}}
\providecommand\npb[3]{{Nucl.\ Phys.\ {\bf B{#1}}, {#3} ({#2})}}
\providecommand\jhep[3]{{J.\ High Energy Phys.\ {\bf #1}, {#3} ({#2})}}
\providecommand\npps[3]{{Nucl.\ Phys.\ {\bf {#1}} Proc.\ Suppl.\ {#3} ({#2})}}

\newcommand\citeprd[3]{{\ifuseprd{Phys.\ Rev.\ D {\bf {#1}}, {#3} ({#2})}%
                        \else{\prd{#1}{#2}{#3}}\fi}}
\newcommand\citeprl[3]{{\ifuseprd{Phys.\ Lett.\ {\bf {#1}}, {#3} ({#2})}%
                        \else{\prl{#1}{#2}{#3}}\fi}}
\providecommand\hepth[1]{{\ifuseprd{\eprint{{\ifeprint\tt\fi hep-th/#1}}}%
                \else{\tt hep-th/{#1}}\fi}}

\newcommand\phepth[1]{{\ifuseprd\else\tt\fi [\hepth{#1}]}}

\newcommand\ki[1]{{\ensuremath{k_{{#1}}}}}
\newcommand\ul[1]{{\breve{{#1}}}}
\newcommand\covd{{\ensuremath{{\mathcal D}}}}
\newcommand\fr[1]{{\ensuremath{{\widehat{#1}}}}}
\newcommand\frfrqw{{\ensuremath{{\fr{\frac{qw}{2}}}}}}

\ifuseprd
\begin{document} 
\fi 

\title{(Twisted) Toroidal Compactification of pp-Waves}
\ifuseprd
\author{Jeremy Michelson}\email{jeremy@physics.rutgers.edu}
\affiliation{New High Energy Theory Center,
       Rutgers University;
       126 Frelinghuysen Road;
       Piscataway, NJ \ 08854}
\else 
\author{
Jeremy Michelson\thanks{\tt jeremy@physics.rutgers.edu} \\
New High Energy Theory Center \\
Rutgers University \\
126 Frelinghuysen Road \\
Piscataway, NJ \ 08854 \ USA 
} 
\fi 

\putabstract{
The maximally supersymmetric type IIB
pp-wave is compactified on spatial circles, with and without an
auxiliary rotational twist.  
All spatial circles of constant radius are identified.
Without the twist, an S$^1$ compactification
can preserve 24, 20 or 16 supercharges.  
$T^2$ compactifications can preserve 20, 18 or 16 supercharges; $T^3$
compactifications can preserve 18 or 16 supercharges and higher
compactifications preserve 16 supercharges.
The worldsheet theory of
this background is discussed.
The T-dual and decompactified
type IIA and M-theoretic solutions which preserve 24
supercharges
are given.
Some comments are made regarding the
AdS parent and the CFT description.
}

\preprint{RUNHETC-2002-08\ifuseprd,~\else\\\fi 
   {\tt hep-th/0203140}}

\ifuseprd
\maketitle
\else
\begin{document}
\fi 

\tableofcontents

\section{Introduction} \label{sec:intro}

In addition to the maximally symmetric AdS$_5\times$S$^5$ and Minkowski
backgrounds, it has recently been realized that there is an additional
background for the type IIB string which preserves 32 supercharges,
namely the pp-wave.\cite{bfhp}
As this background is achieved as a Penrose limit of
AdS$_5\times$S$^5$~\cite{bfhp2,bfp}, string theory on this background has a CFT
dual~\cite{bmn}.
Remarkably, this CFT is powerful enough to see the perturbative string
spectrum.
These observations have caused a flurry of
interest~\cite{m,mt,bhk,ikm,go,rt,pzs,hks,clp,as,bp,kprs,tt,fk,gns}.

Moreover, the pp-wave has 30 isometries---the same
number as AdS$_5\times$S$^5$ by virtue of the Penrose limit---of which
sixteen are spatial%
\iftoomuchdetail
---that is, the Killing vector has positive norm%
\fi\ 
and noncompact.
These therefore provide a way to compactify the pp-wave to lower dimensions.
In fact,
the maximal number of commuting, noncompact and spatial Killing vectors is
eight, the same number as for toroidal compactification of ordinary Minkowski
space to the light cone.

The study of toroidal 
compactifications 
has provided a very
rich structure of phenomena and dualities.  
In this paper such a study is initiated for
pp-waves.  The first step, given after a review of the pp-wave
geometry in section~\ref{sec:geom}, is to identify spacelike
isometries on which to compactify.  This is done in
section~\ref{sec:compact}.

An analysis of Killing spinors, in section~\ref{sec:compact}, shows that
compactification preserves at least half the supersymmetries, up
to $3/4$, or 24 supercharges.  While surprising, this can occur because of the
reduced rotational symmetry of the system, due to the nontrivial
curvatures.
In~\cite{gh} an analysis of the central charge matrix gave rise to the
possibility of $3/4$ BPS states; it would be interesting if
the geometry studied
here is employing this mechanism.  
\skipthis{In particular,
\skipthis{
one would not expect the central charge matrix to be affected by what
is simply a
discrete orbifolding.}
the pp-wave background is not asymptotically flat so it is
not clear what the proper definition is of the central charges of this
background.}%

In section~\ref{sec:Tdual}, 
the T-dual solution is given and lifted to M-theory.  It is then shown
that the M-theory solution also preserves 24 supercharges.  This is
because the Killing spinors on the IIB side do not involve the compact
coordinate.  Otherwise there would be ``supersymmetry without
supersymmetry''~\cite{duff,duff2}.

Since the
isometries derive from AdS$_5\times$S$^5$ isometries, it is possible to
find the corresponding quotient of AdS$_5$, which has a CFT
dual.  As discussed in section~\ref{sec:AdS}, the quotient turns out
to be one of those previously discussed
by Ghosh and Mukhi~\cite{gm}.

In addition we can perform a Melvin-like twist, as was previously done
in the flat space context in many papers,
including~\cite{dgkt,dggh1,dggh2,rtmel1,%
tmel1,tmelrev,rtmel2,tmel,gc,dgmm}.
This breaks all the supersymmetries.  In flat space, the result was
a tachyon in the spectrum.  Understanding this closed string tachyon
is a very interesting problem which has been addressed
in e.g.~\cite{gs,rtt1,aps,tu,rtt2,hkmm,dv,my,s,dgmm,af}.
A closed string tachyon is similarly expected here, the
difference being that there should also be a dual CFT description.
This could be a good way to get some control over closed string
tachyon condensation.

The paper is organized as follows.  In section~\ref{sec:geom} the
supersymmetric type IIB pp-wave is reviewed with an emphasis on the
symmetries, their algebra
and the introduction of a convenient coordinate system for use in most
of the paper.  These results are used in
section~\ref{sec:compact} to identify spacelike Killing vectors and
study the supersymmetries preserved by compactification along their
orbits.  The Green-Schwarz string is quantized on these geometries in
section~\ref{sec:gs}.  The SO(8) symmetry of the metric means that
the quantization of the bosonic part of the string is independent of
the choice of spacelike Killing vector.  For the fermionic part of the
string, we analyse the maximally supersymmetric compactification in
section~\ref{sec:gsferm12} and the minimally (half) supersymmetric
S$^1$ compactification in section~\ref{sec:gsferm15}.
In the latter case, the Hamiltonian is time dependent.
The effect on the compactification of
adding a twist is analysed in section~\ref{sec:Melvin}.  Then
T-duality and supersymmetry is discussed in section~\ref{sec:Tdual}.
Finally the AdS$_5\times$S$^5$ parent orbifold, and the choices
therein, is described in section~\ref{sec:AdS}.
Section~\ref{sec:conc} contains some conclusions.
Two appendices contain some additional useful formulas.
Although most of the text is devoted to compactification on a single
circle, in many cases, the results generalize to higher tori in an
obvious way.  One exception is a compactification on a $T^2$ involving
the same two coordinates (in the ``standard'' coordinate system
of~\cite{bfhp}).
Appendix~\ref{sec:spm} presents a (singular) coordinate system adapted to
this compactification.
In
Appendix~\ref{sec:altkillspin} some additional expressions for the
Killing spinors are given.

\section{The pp-wave} \label{sec:geom}

In this section some useful facts about the pp-wave
geometry, its isometries and its Killing spinors are collected.

The supergravity solution is~\cite{bfhp}
\begin{subequations} \label{bfhp:pp}
\begin{align} 
ds^2 &= 2 dx^+ dx^- - 4\mu^2 x^i x^i (dx^+)^2 + dx^i dx^i, \\
\rrf &= \mu\, dx^+ (dx^1 dx^2 dx^3 dx^4 + dx^5 dx^6 dx^7 dx^8),
\end{align}
\end{subequations}
where $i=1\dots 8$, 
the constant $\mu\neq0$ will be kept arbitrary although
it can be set to any convenient value by a coordinate transformation,
and $\rrf$ is the self dual \RR\ five-form
field strength.  All other fields vanish.
These satisfy the IIB equations of motion, 
including
\begin{equation}
G_{\mu\nu} = \frac{4}{3} F_{\mu\alpha\beta\gamma\delta} 
F_\nu{^{\alpha\beta\gamma\delta}}.
\end{equation}
Greek indices run over all the coordinates.

It will be convenient to make the change of coordinates,
\begin{subequations} \label{myX}
\begin{gather}
\begin{align}
x^+ &= X^+, &
x^- &= X^- - 2 \mu X^1 X^2, &
x^I &= X^I, I= 3\dots 8 
\end{align} \\
\begin{align}
x^1 &= X^1 \cos(2\mu X^+) - X^2 \sin(2\mu X^+), &
x^2 &= X^1 \sin(2\mu X^+) + X^2 \cos(2\mu X^+),
\end{align}
\end{gather}
\end{subequations}
Then the metric takes the form
\begin{subequations} \label{mypp12}
\begin{align}
ds^2 &= 2 dX^+ dX^- - 4\mu^2 X^I X^I (dX^+)^2 - 8 \mu X^2 dX^1 dX^+
+ dX^i dX^i, \\
\rrf &= \mu\, dX^+ (dX^1 dX^2 dX^3 dX^4 + dX^5 dX^6 dX^7 dX^8).
\end{align}
\end{subequations}
In this coordinate system, $\p_1 = \frac{\p}{\p X^1}$ is a manifest
isometry.  This is one of the circles on which we will consider
compactification.  
I%
n this coordinate system, $X^1$, $X^2$ are no
longer massive bosons but instead have the quantum mechanics of a
particle in a plane with constant magnetic field.

\subsection{Isometries and Killing Vectors} \label{sec:iso}

There is an obvious SO(4)$\times$SO(4)$\rtimes\field{Z}_2$ symmetry
which
rotates and exchanges the $\{x^1,x^2,x^3,x^4\}$ and
$\{x^5,x^6,x^7,x^8\}$ subspaces.  In fact, 30 Killing vectors were
identified in~\cite{bfhp}, namely
\begin{subequations} \label{bfhp:kill}
\begin{gather}
\begin{align} 
\ki{e_+} &= -\p_+, & \ki{e_-} &= -\p_-,
\end{align} \\ \label{bfhp:killi}
\ki{e_i} = -\cos(2\mu x^+) \p_i - 2 \mu \sin(2\mu x^+) x^i \p_-, \\
\label{bfhp:killi*}
\ki{e^*_i} = -2\mu \sin(2\mu x^+) \p_i + 4 \mu^2 \cos(2\mu x^+) x^i \p_-, \\
\label{bfhp:killM}
\ki{M_{ij}} = x^i\p_j - x^j \p_i, 
   \text{both $i,j$=$1\dots4$ or both $i,j$=$5\dots8$}.
\end{gather}
\end{subequations}
These obey the algebra
\begin{subequations} \label{bfhp:killalg}
\begin{gather}
\begin{align}
\com{\ki{e_-}}{\text{all}} &= 0, &
&&
\com{\ki{e_+}}{M_{ij}} &= 0,
\\
\com{\ki{e_+}}{\ki{e_i}} &= \ki{e_i^*}, &
\com{\ki{e_+}}{\ki{e_i^*}} &= -4 \mu^2 \ki{e_i}, &
\com{\ki{e_i}}{\ki{e_j^*}} &= 4\mu^2 \delta_{ij} \ki{e_-}, 
\\
\com{\ki{M_{ij}}}{\ki{e_k}} &= \delta_{jk} \ki{e_i} - \delta_{ik} \ki{e_j}, &
&&
\com{\ki{M_{ij}}}{\ki{e_k^*}} 
    &= \delta_{jk} \ki{e_i^*} - \delta_{ik} \ki{e_j^*},
\end{align} \\ 
\com{\ki{M_{ij}}}{\ki{M_{kl}}} = \delta_{jk} \ki{M_{il}} 
- \delta_{ik} \ki{M_{jl}} - \delta_{jl} \ki{M_{ik}} + \delta_{il} \ki{M_{jk}}.
\end{gather}
\end{subequations}
In particular, $M_{ij}$ are the SO(4)$\times$SO(4) rotational
generators.

\subsection{Killing Spinors and the Superalgebra} \label{sec:ks}

Before listing the Killing spinors,
some notation must be introduced.  The ten dimensional
$\Gamma$-matrices obey 
\begin{equation}
\anti{\Gamma^{\ul\mu}}{\Gamma^{\ul\nu}}=2\eta^{\ul\mu\ul\nu},
\qquad \Gamma^{\ul\pm} = \frac{1}{\sqrt{2}} (\Gamma^{\ul9} \pm \Gamma^{\ul0}),
\end{equation}
where $\eta^{\ul\mu\ul\nu}$ is the mostly positive Minkowski metric of the
tangent bundle.
$\Gamma^{\mu_1\dots\mu_p}$ are products of the $\Gamma$-matrices,
antisymmetrized with unit weight.
The zenbein is~\cite{bfhp}
\begin{align}
e^{\ul-} &= dx^- -2 \mu^2 (x^i)^2 dx^+, &
e^{\ul+} &= dx^+, &
e^{\ul{\imath}} &= dx^i.
\end{align}
I have used the cup to emphasize the tangent-space labels; this is
important as
the following formulas employ this choice of local frame.
It is convenient to define
\begin{align}
I &= \Gamma^{\ul1} \Gamma^{\ul2} \Gamma^{\ul3} \Gamma^{\ul4}, &
J &= \Gamma^{\ul5} \Gamma^{\ul6} \Gamma^{\ul7} \Gamma^{\ul8}.
\end{align}
In terms of the one-form
$\Omega_\mu=\frac{i}{24} F_{\mu \alpha \beta \gamma \delta}
                    \Gamma^{\alpha\beta\gamma\delta}$,
\begin{align} \label{defOmega}
\Omega_- &= 0, &
\Omega_+ &= \mu (I + J), &
\Omega_i &= \begin{cases} -\mu \Gamma^{\ul+} \Gamma^{\ul{\imath}} I,
      & i=1\dots 4, \\
                          -\mu \Gamma^{\ul+} \Gamma^{\ul{\imath}} J,
      & i=5\dots 8,
            \end{cases}
\end{align}
to each constant, complex positive chirality spinor $\psi$ is
associated the Killing
spinor~\cite{bfhp}
\begin{equation}
\epsilon(\psi) = \left[1-i x^i \Omega_i\right]
\left[\cos(\mu x^+) \one - i \sin(\mu x^+) I\right]
\left[\cos(\mu x^+) \one - i \sin(\mu x^+) J\right] \psi.
\end{equation}
These obey the Killing spinor equation
\begin{equation} \label{bfhp:killeq}
(\nabla_\mu + \frac{i}{24} F_{\mu \alpha \beta \gamma \delta}
                    \Gamma^{\alpha\beta\gamma\delta})
\epsilon(\psi) = 0.
\end{equation}
Since there are 32 linearly independent $\psi$s, there are 32 linearly
independent positive chirality Killing spinors, corresponding to
${\mathcal N}=2B$ supersymmetry in ten dimensions.


These are acted on nontrivially by the Killing vectors.  Specifically,
the Lie derivative on spinors is defined
by (see~\cite{fp} and references therein)
\begin{equation} \label{liespin}
\lie{\ki{}}{\epsilon} = \ki{}^\mu \nabla_\mu \epsilon + 
\frac{1}{4} (\nabla_\mu \ki{}_\nu) \Gamma^{\mu\nu} \epsilon.
\end{equation}
Then,~\cite{bfhp,m}%
\begin{subequations} \label{bfhp:lieks}
\begin{gather}
\begin{align}
\lie{\ki{e_-}}{\epsilon(\psi)} &= 0, &
\lie{\ki{e_+}}{\epsilon(\psi)} &= \epsilon(i\mu(I+J) \psi), &
\lie{M_{ij}}{\epsilon(\psi)} &= 
    \epsilon(\half \Gamma^{\ul\imath\ul\jmath}\psi),
\end{align} \\ \begin{align}
\lie{\ki{e_{i=1\dots4}}}{\epsilon(\psi)} 
  &= \epsilon(-i\mu I \Gamma^{\ul\imath} \Gamma^{\ul+} \psi) &
\lie{\ki{e_{i=1\dots4}^*}}{\epsilon(\psi)} 
  &= \epsilon(2\mu^2 \Gamma^{\ul\imath} \Gamma^{\ul+} \psi) \\
\lie{\ki{e_{i=5\dots8}}}{\epsilon(\psi)} 
  &= \epsilon(-i\mu J \Gamma^{\ul\imath} \Gamma^{\ul+} \psi) &
\lie{\ki{e_{i=5\dots8}^*}}{\epsilon(\psi)} 
  &= \epsilon(2\mu^2 \Gamma^{\ul\imath} \Gamma^{\ul+} \psi).
\end{align}
\end{gather}
\end{subequations}
This information will be used to count the number of supersymmetries
preserved by compactification.

\section{S$^1$ and $T^d$ Compactification of the pp-wave} \label{sec:compact}

In~\cite{bfhp}, the authors considered compactification along the
circle generated by the [manifest in equation~\eqref{bfhp:pp}]
isometry $\ki{e_-}+\ki{e_+}$.%
\footnote{This was also one of the two isometries that was
compactified in~\cite{rt}, in the
context of the Penrose limit of AdS$_3\times$S$^3\times T^4$.  The
other isometry considered in~\cite{rt} is one associated with the $T^4$.}
However, that is not spacelike:
\begin{equation} \label{bfhp:S1}
\norm{\ki{e_-}+\ki{e_+}}^2 = 2 - 4 \mu^2 (x^i)^2.
\end{equation}
Though compactification of
timelike circles has been considered in~\cite{hull}, for example,
such
a circle is likely to introduce complications 
that I do not want to consider here.%
\footnote{After much of this paper was written,~\cite{clp} appeared which
discusses this type of compactification with
interesting results.}
Also, the isometry~\eqref{bfhp:S1} 
breaks all the supersymmetries, except at special values of the radius.
Instead, there are several isometries that, while not yet manifest,
are spacelike of unit norm.  Namely, define
\begin{equation} \label{mykill}
\ki{S^\pm_{ij}} = \ki{e_i}\pm \frac{1}{2\mu} \ki{e_j^*},
\end{equation}
and note that
\begin{equation}
\norm{\ki{S^\pm_{ij}}}^2 = 1, i\neq j.
\end{equation}
We will use this set of isometries to compactify the pp-wave geometry.
(For fixed $i$, 
\hbox{$\norm{\ki{S^\pm_{ii}}}^2 = 1 \pm \sin 4\mu x^+$} goes null at
$x^+=\frac{(2n\mp1)\pi}{8\mu}$.)
Note that, in contrast to toroidal compactification
of flat space, the isometries~\eqref{mykill}
are not hypersurface orthogonal.

Although there are many of these isometries, 
the SO(4)$\times$SO(4)$\rtimes\field{Z}_2$ symmetry%
\footnote{Including the $\ZZ_2\in SO(4)$ element
$x^i,x^j \rightarrow -x^i,-x^j$ (with $i,j=1\dots4$ or
$i,j=5\dots8$) which takes $S^\pm_{ij}\rightarrow -S^\pm_{ij}$,
$S^\pm_{ik}\rightarrow -S^\mp_{ik}$ and $S^\pm_{kj}\rightarrow
S^\mp_{kj}$, $k\neq i,j$.}
implies that we can, without loss of generality, choose the ``$+$''
sign and $i=1$; then there are only two distinct choices of $j$,
namely $j=2$ or $j=5$.  Although the metric behaviour is the same for both
choices, the \RR\ field is quite different.

For the first choice, $j=2$, the isometry $\ki{S^+_{12}}$ is manifest in
equation~\eqref{mypp12}.  Also,
\begin{equation} \label{x+kill}
\frac{\p}{\p X^+} = -\ki{e_+} + 2 \mu \ki{M_{12}}
\end{equation}
is a Killing vector of the geometry.
For $j=5$%
\
the \RR\
field is more complicated in manifest coordinates:
\begin{subequations} \label{mypp15}
\begin{align}
ds^2 &= 2 dX^+ dX^- - 4\mu^2 X^{\hat{I}} X^{\hat{I}} (dX^+)^2 
- 8 \mu X^5 dX^1 dX^+
+ dX^i dX^i, \\
\begin{split} \label{myf15}
\rrf &= \mu dX^+ [\cos(2\mu X^+) dX^1 dX^2 dX^3 dX^4  
 -\sin(2\mu X^+) dX^5 dX^2 dX^3 dX^4 \\
& \quad  +\sin(2\mu X^+) dX^1 dX^6 dX^7 dX^8
 + \cos(2\mu X^+) dX^5 dX^6 dX^7 dX^8],
\end{split}
\end{align}
\end{subequations}
where $\hat{I}=2,3,4,6,7,8$.  In particular note that
\begin{equation}
\frac{\p}{\p X^+} = \frac{\p}{\p x^+} 
   + 2\mu \left(x^1 \frac{\p}{\p x^5}-x^5 \frac{\p}{\p x^1}\right)
\end{equation}
is not an isometry of the field configuration.
For this choice of circle compactification, then, the light-cone
Hamiltonian will be time dependent.

Focusing on the $S^+_{12}$ compactification, the
nine dimensional field configuration is easily read off~\cite{ms} as
\begin{subequations} \label{S12d9}
\begin{align}
ds_9^2 &= 2 dX^+ dX^- - 4\mu^2 [4(X^2)^2+X^{\hat{I}} X^{\hat{I}}] (dX^+)^2 
+ dX^{\hat{\imath}} dX^{\hat{\imath}}, \\
A_9 &= -4\mu X^2 dX^+, \\
{^{(3)}}A_9 &= -\mu X^+ dX^2 dX^3 dX^4,
\end{align}
\end{subequations}
where $\hat{\imath}=2\dots8$, 
$A_9$ is the KK gauge field and
${^{(3\iftoomuchdetail,4\fi)}}A_9$ \iftoomuchdetail are\else is\fi\ the 
dimensionally reduced potential\iftoomuchdetail s\fi\ for the
ten dimensional self dual
five-form field strength.
\iftoomuchdetail
Self duality of the ten dimensional 
five-form means that we only need to keep ${^{(3)}}A_9$.
\fi

All of the $S^\pm_{ij}$ compactifications {\em preserve\/} at least
$1/2$ of the supersymmetry, or
16 supercharges.  If $i,j$ are in the same $SO(4)$ subgroup, then $3/4$,
or 24 supercharges, is preserved.  This is seen from
equations~\eqref{bfhp:lieks},
since
\begin{equation}
\lie{\ki{e_i}+\frac{1}{2\mu}\ki{e_j^*}}{\epsilon(\psi)}
= 
\epsilon(i\mu Q^{ij} \psi), 
\end{equation}
where
\begin{equation} \label{defQ}
Q^{ij} = \begin{cases} 
\Gamma^{\ul\jmath} (\half \sum_{k,l=1}^4 \epsilon_{ijkl} \Gamma^{\ul k\ul l}
-i \one) \Gamma^{\ul +}, & i,j=1\dots4\text, \\
\Gamma^{\ul\jmath} (\half \sum_{k,l=5}^8 \epsilon_{ijkl} \Gamma^{\ul k\ul l}
-i \one) \Gamma^{\ul+}, & i,j=5\dots8\text, \\
-\Gamma^{\ul\jmath} (\frac{1}{6} \sum_{k,l,m=1}^4
    \epsilon_{iklm} \Gamma^{\ul k\ul l\ul m}
    \Gamma^{\ul\jmath}
+i \one) \Gamma^{\ul+}, & i=1\dots4,j=5\dots8, \\
-\Gamma^{\ul\jmath} (\frac{1}{6} \sum_{k,l,m=5}^8 
    \epsilon_{iklm} \Gamma^{\ul k\ul l\ul m}
    \Gamma^{\ul\jmath}
+i \one) \Gamma^{\ul+}, & i=5\dots8,j=1\dots4,
  \end{cases}
\end{equation}
For the periodic spin structure, the supersymmetries preserved by the
compactification are those whose
Killing spinors are preserved by the isometry.  That is, we want to
count the number of complex, constant, positive chirality spinors
annihilated by $Q$.
[In
principle, we should require only $e^{2\pi i R
\lie{\ki{}}{}} \epsilon(\psi)=\epsilon(\psi)$; however, since $Q^{ij}$
is nilpotent,
this is the same thing.  In particular,
there are no special radii of enhanced supersymmetry.]
Clearly, these include the 16 spinors annihilated by $\Gamma^{\ul+}$.  For
$i,j$ in different $SO(4)$ subgroups---the last two lines
of equation~\eqref{defQ}---the matrix in parenthesis is nondegenerate,
so there are no other supersymmetries.  When $i,j$ are in the same
$SO(4)$ grouping, then the matrix in parenthesis annihilates an
additional 8 spinors.
In this latter case,
\begin{equation} \label{susyproj}
P^{ij} =\proj_+ - \frac{1}{8} {Q^{ij}}^\dagger {Q^{ij}} \proj_+, 
\quad \text{(no sum; $i,j$ in the same SO(4) grouping)},
\end{equation}
is the projection
operator onto
these 24 spinors, where $\proj_+$ is the projection onto positive
chirality
spinors.  
\skipthis{
Taking the trace (and recalling that our ${\mathcal N}=2B$
conventions are such that the spinors are complex) shows that the
compactification breaks 8 supercharges and preserves 24.
}

More generally, we can consider other linear combinations
of Killing vectors.  However,
from the previous analysis it is easy to see that generic linear
combinations of $\ki{e_i}$ and $\ki{e^*_j}$ 
will break all the supersymmetries.  Although the
addition of any multiple of $\ki{e_-}$ to, say, $\ki{S^+_{12}}$ does
not affect the norm---that is, $\ki{S^+_{12}}+\alpha \ki{e_-}$
provides a nice spacelike isometry---such an
addition can be removed via a coordinate transformation generated by
$\ki{S^-_{21}}$; thus there is no need to consider this linear
combination.  Including $\ki{e_+}$ or $\ki{M_{ij}}$ would generically
break all
the supersymmetries (though we will consider the latter in
section~\ref{sec:Melvin}).  This leaves one interesting possibility:
\begin{equation} \label{neatkill}
\alpha \ki{S^+_{12}} + \beta \ki{S^+_{56}}, \quad \alpha^2+\beta^2=1,
\end{equation}
has unit norm and preserves 20 supercharges, namely those preserved by
the projection
$P^{12} P^{56}$.
\iftoomuchdetail
This follows from brute-force computation of the dimension of the
range of $\alpha Q^{12}+\beta Q^{56}$; alternatively, note that if
$\alpha \Gamma^{\ul2} (\Gamma^{\ul3\ul4}-i)\psi = -\beta \Gamma^{\ul6}
(\Gamma^{\ul7\ul8}-i)\psi$, then we can multiply both sides by the
projection operator $(\Gamma^{\ul3\ul4}-i)(\Gamma^{\ul7\ul8}-i)$.
Since $\alpha \Gamma^{ul2}+\beta
\Gamma^{\ul6}$ is nondegenerate, This tells us that $\psi$ must be
in the kernel of the above projection operator; i.e. annihilated by at
least one side of the equation.  Thus it is annihilated by both
sides; i.e. preserved precisely by $P^{12} P^{56}$.
\fi

Finally, note that if we compactify simultaneously on the circles
generated by both
$\ki{S^+_{ij}}$ and $\ki{S^-_{ij}}$, then we break half the
supersymmetry; namely the half not annihilated by $\Gamma^{\ul+}$.
This is also true for compactification with respect to both
$\ki{S^+_{ij}}$ and
$\ki{S^+_{ji}}$.  For simultaneous compactification on, say, $\ki{S^+_{ij}}$
and $\ki{S^+_{ik}}$, $k\neq j$, again half the supersymmetry is broken.%
\skipthis{\footnote{In this case, the projection operators~\eqref{susyproj} do
not commute; however we can still ask if there are any spinors $\psi$ for
which both $P^{ij} \psi = \psi$ and $P^{ik}\psi=\psi$.  This is
obviously satisfied by $\psi$ in the kernel of $\Gamma^{\ul+}$; otherwise, a
necessary condition, from $(P^{ij}-P^{ik})\psi = 0$,
is that $\Gamma^{\ul+} \proj_+ \psi$ be in the (empty) kernel
of the (nondegenerate)
$\Gamma^{\ul\imath}(\Gamma^{\ul\jmath}-\Gamma^{\ul k})$.}
} 
Although
$\ki{S^-_{ji}}$ preserves the same supersymmetries as $\ki{S^+_{ij}}$,
the two do not commute, and so the compactification on both
does not correspond to a $T^2$
compactification.
More specifically, in the coordinates~\eqref{myX}, if $S^+_{ij}$ is
compactified on a circle of radius $R_1$
and $S^-_{ji}$ is compactified on a circle of radius $R_2$, then we
must also identify the null coordinate
$X^-\sim X^-+8\pi \mu R_1 R_2$.

Thus, we can compactify on at most three circles before we have, at
most, sixteen supercharges.
The coordinate transformation which manifests the
circles is
an obvious generalization of equation~\eqref{myX}, and in this
coordinate system we have obvious generalizations of~\eqref{mypp12}
and~\eqref{mypp15}.

The compactification also breaks some bosonic symmetry.  For
compactification on $S^+_{12}$, the SO(4)$\times$SO(4) is broken to
SO(2)$\times$SO(4), and for the compactification on $S^+_{15}$, the
residual rotational symmetry is SO(3)$\times$SO(3).  Also, 
though $\ki{e_+}$,
$\ki{e_j}$, $\ki{e_i^*}$ are no longer isometries after
compactification on the orbit of $S^+_{ij}$, the linear combinations
$\ki{e_+}-2\mu \ki{M_{12}}$ and
$\ki{e_j}+\frac{1}{2\mu}\ki{e_i^*}$ are still isometries.  
Thus, the residual
isometry group is 23 [$g\rtimes$(SO(3)$\times$SO(3))] or 24-dimensional
[$g\rtimes$(SO(2)$\times$SO(4))] where $g$ is the 17-dimensional 
group
generated by the noncompact Killing vectors $\ki{e_-}$, 
$\ki{e_+}-2\mu \ki{M_{12}}$,
$S^+_{ji}$,
$e_{k\neq j}$, $e_{k\neq i}^*$.  For each additional compactified
direction, 
the rotational group decreases further (to what depends on the details
of the compactification; see table~\ref{tab:sym}), but only removes one
additional generator from
$g$.

\TABLE{\squeezetable
\begin{tabular}{||c|c|c|c|c||}
\toprule
\myparbox[b]{Compactified}{Dimensions Compactified}
& Isometries & \myparbox[b]{Continuous}{Continuous Rotation\skipthis{al
    Symmetry Group}s} & 
    \myparbox[b]{Bosonic Generators}{Number of Other Bosonic Generators}&
    \myparbox[b]{Supercharges}{Preserved Supercharges}
  \\
\colrule
  & $S^+_{12}$ & SO(2)$\times$SO(4) & &24 \\
1 & $\alpha S^+_{12}+\beta S^+_{56}$ & SO(2)$\times$SO(2) & 17 & 20 \\
  & $S^+_{15}$ & SO(3)$\times$SO(3) & &16 \\
\colrule
  & $S^+_{12}, S^+_{34}$ & SO(4) & & 20 \\ 
  & $S^+_{12}, S^+_{56}$ & SO(2)$\times$SO(2) & & 20 \\
  & $\alpha S^+_{12}+\beta S^+_{56}$, $S^+_{34}$ & SO(2) & & 18 \\
  & $S^+_{12}, S^+_{35}$ & SO(3)              & & 16 \\
  & $S^+_{15}, S^+_{26}$ & SO(2)$\times$SO(2) & & 16 \\ 
  & $S^+_{12}, S^+_{21}$ & SO(2)$\times$SO(4) & & 16 \\
2 & $S^+_{12}, S^-_{12}$ & SO(2)$\times$SO(4) & 16 & 16 \\
  & $S^+_{12}, S^+_{13}$ & SO(4) & & 16 \\ 
  & $S^+_{12}, S^+_{15}$ & SO(2)$\times$SO(3) & & 16 \\
  & $S^+_{15}, S^+_{51}$ & SO(3)$\times$SO(3) & & 16 \\
  & $S^+_{15}, S^-_{15}$ & SO(3)$\times$SO(3) & & 16 \\
\colrule
  & $S^+_{12}, S^+_{34}, S^+_{56}$ & SO(2) & & 18 \\ 
  & $S^+_{12}, S^+_{35}, S^+_{46}$ & SO(2) & & 16 \\ 
  & $S^+_{12}, S^+_{35}, S^+_{67}$ & ---   & & 16 \\
  & $S^+_{15}, S^+_{26}, S^+_{37}$ & ---   & & 16 \\
  & $S^+_{12}, S^+_{13}, S^+_{56}$ & SO(2) & & 16 \\
  & $S^+_{12}, S^+_{13}, S^+_{45}$ & SO(3) & & 16 \\
  & $S^+_{12}, S^+_{32}, S^+_{56}$ & SO(2) & & 16 \\
3 & $S^+_{12}, S^+_{32}, S^+_{45}$ & SO(3) & 15 & 16 \\
  & $S^+_{12}, S^+_{32}, S^+_{34}$ & SO(4) & & 16 \\
  & $S^+_{12}, S^-_{12}, S^+_{34}$ & SO(4) & & 16 \\
  & $S^+_{12}, S^-_{12}, S^+_{56}$ & SO(2)$\times$SO(2) & & 16 \\
  & $S^+_{12}, S^-_{12}, S^+_{35}$ & SO(3) & & 16 \\
  & $S^+_{12}, S^+_{35}, S^-_{35}$ & SO(3) & & 16 \\
  & $S^+_{12}, S^+_{21}, S^+_{34}$ & SO(4) & & 16 \\
  & $S^+_{12}, S^+_{21}, S^+_{56}$ & SO(2)$\times$SO(2) & & 16 \\
  & $S^+_{12}, S^+_{21}, S^+_{35}$ & SO(3) & & 16 \\
  & $S^+_{12}, S^+_{35}, S^+_{53}$ & SO(3) & & 16 \\
\colrule
$d\geq$4 & many choices & various & $18-d$ & 16 \\
\botrule
\end{tabular}
\caption{A summary of 
(many of)
the independent ways in which one can compactify on
commuting isometries, and the corresponding unbroken rotational symmetries
and the number of unbroken supercharges.
Not all compactifications preserving 16 supercharges are listed.
The isometries are labeled
as in equation~\eqref{mykill}.  Only the continuous part of the rotational symmetries is listed.\label{tab:sym}}
} 

This is summarized in Table~\ref{tab:sym}
(on page~\pageref{tab:sym}).

\section{The Green-Schwarz String on the Compactified pp-Wave} \label{sec:gs}

In this section the Green-Schwarz string is quantized on the
compactified pp-wave.  The focus is on the compactification along the
orbit of $\ki{S^+_{12}}$; that is, $X^1\sim X^1+2\pi R$ for the
solution~\eqref{mypp12}.  Section~\ref{sec:gsferm15} will briefly
discuss the $S^+_{15}$ compactification.  The bosons, of course, are
impervious to the difference between $S^+_{12}$ and $S^+_{15}$.

\subsection{Bosonic Oscillations} \label{sec:gsbos}

The worldsheet coordinates are $\tau,\sigma$ and the 
Lorentzian worldsheet metric
is $\gamma_{ab}$.
Light-cone gauge is given%
\footnote{Light-cone gauge is allowed since $X^+$ obeys a harmonic
equation of motion before gauge fixing.}
by setting $X^+=\tau$, $\det \gamma=-1$,
$\p_\sigma \gamma_{\sigma\sigma}=0$ and $\gamma_{\tau\sigma}=0$.
The bosonic action in light-cone gauge is then
\begin{equation} \label{Sbos}
S_{\text{B}} = -\frac{1}{4\pi\apr} 
\int d^2 \sigma
\left\{-\gamma_{\sigma\sigma} 
\left[2 \dot{X}^- - 4 \mu^2 \left(X^I\right)^2
      -8\mu X^2 \dot{X}^1 + \left(\dot{X}^i\right)^2
\right]
+ \gamma_{\sigma\sigma}^{-1} 
\left({X^i}{'}\right)^2
\right\},
\end{equation}
where overdots and primes denote differentiation with respect to
$\tau$ and $\sigma$, respectively.
This is supplemented with the constraint equation, of which the
bosonic part [see also equation~\eqref{fst}] is
\begin{gather} \label{bst}
{X^-}{'} = 4 \mu X^2 {X^1}{'} - \dot{X}^i {X^i}{'}. 
\end{gather}
We see that, as usual, $p_- = \frac{\delta S}{\delta\dot{X}^-} =
\frac{\gamma_{\sigma\sigma}}{2\pi\apr}$, along with the
equation [which also gets a fermionic contribution in~\eqref{fss}]
\begin{equation} \label{bss}
\dot{X}^- = 4 \mu X^2 \dot{X}^1 + 2 \mu^2 X^I X^I
-\half \left(\dot{X}^i\right)^2 
- \frac{1}{2 p_-^2 \apr^2} \left({X^i}{'}\right)^2.
\end{equation}

The equation of motion for $X^I$ is the same massive equation as for
the uncompactified pp-wave~\cite{m,mt}.  For $X=X^1+iX^2$,
\begin{subequations}
\begin{align}
\ddot{X} - c^2 {X}{''} &= -4 \mu i\dot{X}, \\
\Ddot{\bar{X}} - c^2 {\bar{X}}{''} &= 4 \mu i\Dot{\bar{X}},
\end{align}
\end{subequations}
and $X\sim X+2\pi R$, where $c=(\apr p_-)^{-1}$.
Thus the mode expansion 
is
\begin{subequations} \label{bosmod}
\begin{align}
\begin{split} \label{modX}
X &= 
X_0 e^{-2\mu i \tau} \cos 2\mu \tau 
+ \frac{p_0}{2\mu p_-} e^{-2\mu i \tau} \sin 2\mu \tau 
+ w R \sigma
\\ & 
+ \frac{i}{p_- \sqrt{\apr}}
\sum_{n\neq0} \frac{e^{-2\mu i \tau}}{n\sqrt{c^2 +4 \frac{\mu^2}{n^2}}} 
\left[ 
\alpha_n e^{-i n\left(\sqrt{c^2+4\mu^2/n^2}\tau+\sigma\right)}
+ \tilde{\alpha}_n e^{-i n\left(\sqrt{c^2+4\mu^2/n^2}\tau-\sigma\right)}
\right], 
\end{split} \\
\begin{split} \label{modbX}
\bar{X} &= 
\bar{X}_0 e^{2\mu i \tau} \cos 2\mu \tau 
+ \frac{\bar{p}_0}{2\mu p_-} e^{2\mu i \tau} \sin 2\mu \tau 
+ w R \sigma
\\ &
+ \frac{i}{p_- \sqrt{\apr}}
\sum_{n\neq0} \frac{e^{2\mu i \tau}}{n \sqrt{c^2 +4 \frac{\mu^2}{n^2}}} 
\left[
\bar{\alpha}_n e^{-i n\left(\sqrt{c^2+4\mu^2/n^2}\tau+\sigma\right)}
+ \Bar{\tilde{\alpha}}_n e^{-i n\left(\sqrt{c^2+4\mu^2/n^2}\tau-\sigma\right)}
\right],
\end{split} \\ \label{modXI}
\begin{split}
X^I &= X_0^I \cos 2 \mu \tau + \frac{p_0^I}{2\mu p_-} \sin 2 \mu \tau
\\ &
+ \frac{i}{p_- \sqrt{\apr}}
\sum_{n\neq0} \frac{1}{n\sqrt{c^2+4\frac{\mu^2}{n^2}}} \left[
\alpha^I_n
    e^{-i n \left(\sqrt{c^2+4\mu^2/n^2}\tau+\sigma\right)}
+ \tilde{\alpha}^I_n
    e^{-i n \left(\sqrt{c^2+4\mu^2/n^2}\tau-\sigma\right)}
\right].
\end{split}
\end{align}
\end{subequations}
Note that
$\alpha_n$, $\tilde{\alpha}_n, \ldots$ is a positive (negative)
frequency mode for $n$
positive (negative) (and large).
The mode expansion for $X, \bar{X}$ is reminiscent of the mode expansion
for the massive scalar, but is rotated by
an 
extra time dependent phase factor $e^{2\mu i \tau}$.
\skipthis{
This is somewhat deceptive, though, in that $X_0$, $\bar{X}_0$---while
named for the fact that they look like center of mass coordinates for
the string---arise from a combination of position and momenta under
the coordinate transformation~\eqref{myX}.  Indeed, the canonical
commutation relations for the zero modes are quite nonstandard,
}
The canonical commutation relations are
\begin{subequations}
\begin{gather} \label{xx,pp_com}
\begin{align}
\com{X_0}{\bar{p}_0} &= i = \com{\bar{X}_0}{p_0}&
\com{X^I_0}{p^J_0} &= i \delta^{IJ},
\end{align} \\ \begin{align}
\com{\alpha_n}{\bar{\alpha}_m} 
&= 2 n \sqrt{1+4\frac{\mu^2}{n^2 c^2}} \delta_{n,-m}
= \com{\tilde{\alpha}_n}{\Bar{\tilde{\alpha}}_m}, \\
\com{\alpha^I_n}{\alpha^J_m} 
&= n \sqrt{1+4\frac{\mu^2}{n^2 c^2}} \delta^{IJ} \delta_{n,-m}
= \com{\tilde{\alpha}^I_n}{\tilde{\alpha}^J_m}, &
\end{align}
\end{gather}
\end{subequations}
and all others vanish.  
\skipthis{
In particular, $\com{X_0}{\bar{p}_0}=0$, etc.
In fact, $p_{\text{phys}} =\int_0^{2\pi} \Pi(\tau,\sigma) =  
\mu p_- i (X_0-\bar{X}_0) - \frac{i}{2}(\bar{p}_0 e^{4\mu i \tau} 
+ p_0 e^{-4\mu i \tau})$, so the strange commutation
relations~\eqref{xx,pp_com} are a result of poor definition rather
than any special physics.  In fact, $X_0=\half \tilde{X}_0 +
\frac{\tilde{p}_0}{4\mu i p_-}, p_0=\mu
p_-\tilde{X}_0-\frac{\tilde{p}_0}{2i}$, and the complex conjugate
equations, give conventional commutation relations---in particular
$\com{\tilde{X}_0}{\Bar{\tilde{p}}_0}=i$---but still does not give a
nice expression for the $p_{\text{phys}}$.  However, the tilded
variables are natural when $\mu\rightarrow 0$.
}
Finally, note that
\begin{subequations}
\begin{align}
X_0^\dagger &= \bar{X}_0, &
p_0^\dagger &= \bar{p}_0, &
{p^I_0}^\dagger &= p^I_0, &
{X^I_0}^\dagger &= X^I_0, \\
\alpha_n^\dagger &= \bar{\alpha}_{-n}, &
\tilde{\alpha}_n^\dagger &= \Bar{\tilde{\alpha}}_{-n}, &
{\alpha^I_n}^\dagger &= \alpha^I_{-n}, &
{\tilde{\alpha}^I_n}{^\dagger} &= \tilde{\alpha}^I_{-n}.
\end{align}
\end{subequations}

\skipthis{
The Hamiltonian is
\begin{equation}
H = \int_0^{2\pi} d\sigma \left\{ \frac{\pi}{p_-} \left[\Pi^1 \Pi^1 
+ \Pi^2
}

In terms of oscillators, the bosonic part of the Hamiltonian is
(ignoring the zero-point energy, which supersymmetry cancels against
the fermionic zero-point energy)
\begin{multline}
H_{\text{B}} = 
2 \mu^2 p_- X_0 \bar{X}_0
+ 2 \mu^2 p_- (X_0^I)^2 
+ \frac{p_0 \bar{p}_0}{2 p_-}
+ \frac{(p_0^I)^2}{2 p_-}
+ \frac{(wR)^2}{2 \apr^2 p_-}
\\
+ c \sum_{n>0} 
\left(\alpha_{-n} \bar{\alpha}_n + \bar{\alpha}_{-n} \alpha_{n} 
   + \tilde{\alpha}_{-n} \Bar{\tilde{\alpha}}_n
   + \Bar{\tilde{\alpha}}_{-n} \tilde{\alpha}_{n} 
   + 2 \alpha_{-n}^I \alpha_n^I 
   + 2 \tilde{\alpha}_{-n}^I \tilde{\alpha}_n^I \right)
\\
- i \mu (X_0 \bar{p}_0 - \bar{X}_0 p_0)
+ c \sum_{n>0} \frac{2 \mu}{n\sqrt{c^2+4\mu^2/n^2}}
\left(\alpha_{-n} \bar{\alpha}_n - \bar{\alpha}_{-n} \alpha_{n} 
   + \tilde{\alpha}_{-n} \Bar{\tilde{\alpha}}_n 
   - \Bar{\tilde{\alpha}}_{-n} \tilde{\alpha}_{n} \right).
\end{multline}
Note that this is not the same Hamiltonian as in~\cite{m,mt}; in
this coordinate system that is natural for compactification, the
Hamiltonian includes the angular momentum generator in the
12-plane---see equation~\eqref{x+kill}.

\subsection{Fermionic Oscillations} \label{sec:gsferm}

The Type IIB Green-Schwarz string is written in terms of a pair of
positive chirality
Majorana-Weyl
space-time spinors $\Theta^\Lambda$, $\Lambda=1,2$.
Along with the index $\Lambda$ come the matrices
\begin{align}
\rho_0 &= i \sigma^2, &
\rho_1 &= \sigma^1, &
\rho_3 &= \sigma^3,
\end{align}
where the $\sigma^i$ are the Pauli matrices.
The starting point
is the observation that with sufficient symmetry
and the light-cone gauge-fixing
\begin{equation} \label{fermlc}
\Gamma^{\ul+} \Theta^\Lambda = 0,
\end{equation}
one can immediately write
the covariantized action~\cite{mt,chsw,ft} 
\begin{equation} \label{mt:SF}
S_{\text{F}} = -\frac{1}{2\pi}\int d^2 \sigma \, \sqrt{-\gamma} \,
i \left(\gamma^{ab} \delta_{\Lambda\Sigma} +
   2\pi \epsilon^{ab} \rho_{3\Lambda\Sigma}\right) 
\p_a X^\mu \Bar{\Theta}^\Lambda \Gamma_\mu \left(\covd_b
\Theta\right)^\Sigma,
\end{equation}
where, in the Majorana representation
$\bar{\Theta}^\Lambda = (\Theta^\Lambda)^\transpose \Gamma^{\ul0}$, and
the covariant derivative on spinors is
\begin{equation} \label{defD}
\covd_a = \p_a 
+ \frac{1}{4} \p_a X^\mu \omega_{\mu\ul{\sigma}\ul{\tau}} 
  \Gamma^{\ul\sigma\ul\tau}
- \frac{1}{2\cdot 5!} \rrf_{\mu\nu\lambda\sigma\tau} 
  \Gamma^{\mu\nu\lambda\sigma\tau} \p_a X^\rho \Gamma_\rho \rho_0,
\end{equation}
where $\omega_{\mu\ul\sigma\ul\tau}$ is the spacetime spin connection and
I have included the effect of a background five-form
field strength, but no other background fields.  (See e.g.~\cite{mt}
for a more complete expression.)
The worldsheet metric is that of section~\ref{sec:gsbos} and
$\epsilon^{ab}$ is a true tensor with $\epsilon_{\tau\sigma}=\sqrt{-\gamma}=1$.

\subsubsection{Compactification on $S^+_{12}$} \label{sec:gsferm12}

The natural coordinate system for compactification along
$\ki{S^+_{12}}$ is that of equation~\eqref{mypp12}, for which
the zenbein is
\begin{align} \label{myzen}
e^{\ul-} &= dX^- - 2\mu^2 (X^I)^2 dX^+ - 4\mu X^2 dX^1, &
e^{\ul+} &= dX^+, &
e^{\ul\imath} &= dX^i,
\end{align}
and the non-zero components of the spin connection are
\begin{align} \label{myspin}
\omega_{+\ul+\ul I} &= - 4\mu^2 X^I, &
\omega_{1\ul+\ul2} &= -2\mu = -\omega_{2\ul+\ul1} 
=
\omega_{+\ul1\ul2}
.
\end{align}
In particular, note that, for this geometry, 
four-and-higher-Fermi Riemann curvature terms
do not appear in
equation~\eqref{mt:SF}, due to the gauge fixing~\eqref{fermlc}.

Thus for the background~\eqref{mypp12}, the fermionic action is%
\iftoomuchdetail
\footnote{This could be written in a more worldsheet covariant way in
which the spacetime spinors are also worldsheet spinors, but that
would be slightly more obscure.}
\fi
\begin{equation} \label{SF12}
S_{\text{F}} = i\apr p_- \int d^2\sigma \,
\left[
\bar{\Theta}^\Lambda \Gamma^{\ul-} 
   \left(\delta_{\Lambda\Sigma}\p_\tau 
         + c \rho_{3\Lambda\Sigma} \p_\sigma\right)\Theta^\Sigma
- 2\mu \bar{\Theta}^\Lambda \Gamma^{\ul-} I \rho_{0\Lambda\Sigma}\Theta^\Sigma
- \mu \bar{\Theta}^\Lambda \Gamma^{\ul-\ul1\ul2} \Theta^\Lambda
\right].
\end{equation}   
\iftoomuchdetail
In deriving this equation the identity $\Gamma^- I \Theta = \Gamma^-
J \Theta$ on positive chirality spinors has been used.
\fi
This leads to the equations of motion
\begin{subequations} \label{feom12}
\begin{align}
\left(\p_\tau + c \p_\sigma - \mu \Gamma^{\ul1\ul2}\right) \Theta^1
- 2 \mu I \Theta^2 &= 0, \\
\left(\p_\tau - c \p_\sigma - \mu \Gamma^{\ul1\ul2}\right) \Theta^2
+ 2 \mu I \Theta^1 &= 0.
\end{align}
\end{subequations}
Also the constraint equation~\eqref{bst} should be replaced with
\begin{gather} \label{fst}
{X^-}{'} = 4 \mu X^2 {X^1}{'} - \dot{X}^i {X^i}{'}
           - i \apr \bar{\Theta}^\Lambda \Gamma^{\ul-} \p_\sigma
             \Theta^\Lambda,
\end{gather}
and equation~\eqref{bss} gets a fermionic contribution so that it reads
\begin{multline} \label{fss}
\dot{X}^- = 4 \mu X^2 \dot{X}^1 + 2 \mu^2 X^I X^I
-\half \left(\dot{X}^i\right)^2 
- \frac{1}{2 p_-^2 \apr^2} \left({X^i}{'}\right)^2
\\
- i \apr \Bar{\Theta}^\Lambda \Gamma^{\ul-} \p_\tau \Theta^\Lambda
+ i \apr \Bar{\Theta}^\Lambda \Gamma^{\ul-}\Gamma^{\ul1\ul2}\Theta^\Lambda
  + 2 i \apr \Bar{\Theta}^\Lambda \Gamma^{\ul-} I \rho_{0\Lambda\Sigma}
      \Theta^\Sigma.
\end{multline}

The general solution to the equations of motion~\eqref{feom12}, subject to the
periodicity condition
\begin{equation} \label{monodromy}
\Theta^\Lambda(\sigma+2\pi,\tau) = \Theta^\Lambda(\sigma,\tau),
\end{equation}
is
\begin{subequations} \label{fermmodes}
\begin{align}
\begin{split}
\Theta^1 &= 
\sqrt{\frac{c}{2\pi}} 
e^{\mu \Gamma^{\ul1\ul2} \tau} (\cos2\mu\tau) \theta_0^1
+ \sqrt{\frac{c}{2\pi}} 
e^{\mu \Gamma^{\ul1\ul2}\tau} (\sin2\mu\tau) I \theta_0^2
\\ & \quad
+ \sqrt{\frac{c}{2\pi}}
\sum_{n\neq 0} 
\frac{e^{\mu \Gamma^{\ul1\ul2} \tau}}{\sqrt{1+\left(
      \frac{n(\sqrt{c^2+4\mu^2/n^2}-c)}{2\mu}\right)^2}}
\left[ 
e^{-i n\left(\sqrt{c^2+4\mu^2/n^2}\tau-\sigma\right)} \theta_n^1
\right. \\ & \qquad \qquad \left.
+ \frac{i}{2\mu} n \left(\sqrt{c^2+\frac{4\mu^2}{n^2}} - c\right)
   e^{-i n\left(\sqrt{c^2+4\mu^2/n^2}\tau+\sigma\right)} I \theta_n^2
\right], 
\end{split} \\ \begin{split}
\Theta^2 &= 
\sqrt{\frac{c}{2\pi}}
e^{\mu \Gamma^{\ul1\ul2} \tau} (\cos2\mu\tau) \theta_0^2
- \sqrt{\frac{c}{2\pi}} 
e^{\mu \Gamma^{\ul1\ul2}\tau} (\sin2\mu\tau) I \theta_0^1
\\ & \quad
+ \sqrt{\frac{c}{2\pi}} 
\sum_{n\neq 0} 
\frac{e^{\mu \Gamma^{\ul1\ul2} \tau}}{\sqrt{1+\left(
      \frac{n(\sqrt{c^2+4\mu^2/n^2}-c)}{2\mu}\right)^2}}
\left[ 
e^{-i n\left(\sqrt{c^2+4\mu^2/n^2}\tau+\sigma\right)} \theta_n^2
\right. \\ & \qquad \qquad \left.
- \frac{i}{2\mu} n \left(\sqrt{c^2+\frac{4\mu^2}{n^2}} - c\right)
   e^{-i n\left(\sqrt{c^2+4\mu^2/n^2}\tau-\sigma\right)} I \theta_n^1
\right],
\end{split}
\end{align}
\end{subequations}
where $\theta_n^\Lambda$ are positive chirality
Majorana-Weyl
spinor-operators for which
$\Gamma^{\ul+}\theta_n^\Lambda=0$.  
Comparing to~\cite{m,mt}, we see
that the fermions are rotated like the bosons, but in the spin
representation, of course.
Half of the $\theta_n^\Lambda$s (or four for each
$\Lambda$) have $\Gamma^{\ul1\ul2}=i$ and the other half have
$\Gamma^{\ul1\ul2}=-i$.
Note that the periodicity
condition~\eqref{monodromy}
is unaffected by compactification in the $X^1$-direction, again due in
part to the gauge
fixing~\eqref{fermlc}; thus equation~\eqref{fermmodes} holds in
winding sectors, as well.

Reality of $\Theta^\Lambda$ implies
\begin{equation}
{\theta^{\alpha\Lambda}_n}{^\dagger} = \theta^{\alpha\Lambda}_{-n},
\end{equation}
as usual%
\iftoomuchdetail%
(in particular, $\Gamma^{12}$ is real)%
\fi
, where here, $\alpha$ is a spinor index; in particular, the zero
modes are self-conjugate.
The canonical commutation relations are
\begin{equation}
\anti{\theta_n^{\alpha\Lambda}}{\theta_m^{\beta\Sigma}} =
\frac{1}{\sqrt{2}}
\delta_{n,-m} 
\frac{\left(\Gamma^{\ul+}\Gamma^{\ul-}\right)^{\alpha\beta}}{2}
\delta^{\Lambda\Sigma},
\end{equation}
where the $\Gamma$ matrices enforce light-cone gauge by projecting
onto the subspace annihilated by
$\Gamma^{\ul+}$.

The fermionic contribution to the Hamiltonian is
(again ignoring the zero-point energy)
\begin{multline} \label{HF}
H_{\text{F}} =
i \mu \bar{\theta}_0^1 \Gamma^{\ul-\ul1\ul2} \theta_0^1
+i \mu \bar{\theta}_0^2 \Gamma^{\ul-\ul1\ul2} \theta_0^2
+4\mu i \bar{\theta}_0^1 \Gamma^{\ul-} I \theta_0^2
\\
+2 \sum_{n>0} 
\left[\bar{\theta}^1_{-n} \Gamma^{\ul-} 
                 \left(n \sqrt{c^2+\frac{4\mu^2}{n^2}} 
                      +i \mu \Gamma^{\ul1\ul2}\right)\theta^1_n 
      + \bar{\theta}^2_{-n} \Gamma^{\ul-} 
                  \left(n \sqrt{c^2+\frac{4\mu^2}{n^2}} 
                      +i \mu \Gamma^{\ul1\ul2}\right)\theta^2_n\right]
\end{multline}
Again, this is recognizable
as the Hamiltonian of~\cite{m,mt} plus fermionic angular momentum
in the 12-plane.

\subsubsection{Compactification on $S^+_{15}$} \label{sec:gsferm15}

For compactification along the circle generated by $\ki{S^+_{15}}$
the zenbein is
\begin{align} \label{myzen15}
e^{\ul-} &= dX^- - 2\mu^2 (X^I)^2 dX^+ - 4\mu X^5 dX^1, &
e^{\ul+} &= dX^+, &
e^{\ul\imath} &= dX^i,
\end{align}
and more importantly, the field strength~\eqref{myf15} has gained some
$X^+$-dependence.  As a result, the action~\eqref{mt:SF} is now
\begin{multline}
S_{\text{F}} = i\apr p_- \int d^2\sigma \,
\left[
\bar{\Theta}^\Lambda \Gamma^{\ul-} 
   \left(\delta_{\Lambda\Sigma}\p_\tau 
         + c \rho_{3\Lambda\Sigma} \p_\sigma\right)\Theta^\Sigma
\right. \\ \left.
- 2\mu \bar{\Theta}^\Lambda \Gamma^{\ul-} 
  I e^{-2\mu \tau \Gamma^{\ul1\ul5}} \rho_{0\Lambda\Sigma}\Theta^\Sigma
- \mu \bar{\Theta}^\Lambda \Gamma^{\ul-\ul1\ul5} \Theta^\Lambda
\right].
\end{multline}
The action is now time dependent.
Despite this difference, the classical solution to the equations of motion is
still given by the mode expansion~\eqref{fermmodes}, after replacing
$\Gamma^{\ul1\ul2}$ with $\Gamma^{\ul1\ul5}$.
Note that unlike
there, the ordering of the $\Gamma$ matrices is important here since
$\Gamma^{\ul1\ul5}$ {\em anti\/}commutes with $I$.
Explicitly,
\begin{subequations} \label{fermmod15}
\begin{align}
\begin{split}
\Theta^1 &= 
\sqrt{\frac{c}{2\pi}} 
e^{\mu \Gamma^{\ul1\ul5} \tau} (\cos2\mu\tau) \theta_0^1
- e^{\mu \Gamma^{\ul1\ul5} \tau} \sqrt{\frac{c}{2\pi}} 
(\sin2\mu\tau) I \theta_0^2
\\ & \quad
+ \sqrt{\frac{c}{2\pi}}
\sum_{n\neq 0} 
\frac{e^{\mu \Gamma^{\ul1\ul5} \tau}}{\sqrt{1+\left(
      \frac{n(\sqrt{c^2+4\mu^2/n^2}-c)}{2\mu}\right)^2}}
\left[
e^{-i n\left(\sqrt{c^2+4\mu^2/n^2}\tau-\sigma\right)} \theta_n^1
\right. \\ & \qquad \qquad \left.
+ \frac{i}{2\mu} n \left(\sqrt{c^2+\frac{4\mu^2}{n^2}} - c\right)
   e^{-i n\left(\sqrt{c^2+4\mu^2/n^2}\tau+\sigma\right)} I \theta_n^2
\right], 
\end{split} \\ \begin{split}
\Theta^2 &= 
\sqrt{\frac{c}{2\pi}}
e^{\mu \Gamma^{\ul1\ul5} \tau} (\cos2\mu\tau) \theta_0^2
+ \sqrt{\frac{c}{2\pi}}
e^{\mu \Gamma^{\ul1\ul5} \tau} (\sin2\mu\tau) I \theta_0^1
\\ & \quad
+ \sqrt{\frac{c}{2\pi}} 
\sum_{n\neq 0} 
\frac{e^{\mu \Gamma^{\ul1\ul5} \tau}}{\sqrt{1+\left(
      \frac{n(\sqrt{c^2+4\mu^2/n^2}-c)}{2\mu}\right)^2}}
\left[
e^{-i n\left(\sqrt{c^2+4\mu^2/n^2}\tau+\sigma\right)} \theta_n^2
\right. \\ & \qquad \qquad \left.
- \frac{i}{2\mu} n \left(\sqrt{c^2+\frac{4\mu^2}{n^2}} - c\right)
   e^{-i n\left(\sqrt{c^2+4\mu^2/n^2}\tau-\sigma\right)} I \theta_n^1
\right].
\end{split}
\end{align}
\end{subequations}
So, not surprisingly, we again obtain the rotated uncompactified result.

\section{Twisted Compactification} \label{sec:Melvin}

In this section 
the compactification is done along the
isometry
\begin{equation} \label{melisom}
-\ki{S^+_{12}} + \frac{q}{R} \ki{M_{34}}.
\end{equation}
In flat space, this compactification leads to the Melvin
universe---see e.g.~\cite{dgkt,dggh1,dggh2,rtmel1,tmel1,tmelrev,rtmel2,tmel}.%
\footnote{Compared to much of the literature,
$q_{\text{here}}=(qR)_{\text{there}}$.  In particular,
$q_{\text{here}}$ is dimensionless.}
I have chosen to rotate in the $34$-plane as the circle is traversed,
but little will change in the following [though the rotational symmetry
of the resulting space would be SO(2)$\times$SO(2) instead of
SO(4)] if the rotation is, say, in the $56$-plane instead.
Note that---unless $q$ is an even integer, at which point the theory is
equivalent to compactification along $\ki{S^+_{12}}$---all the
supersymmetry is broken by this compactification.

With
\begin{equation} \label{defZ}
Z = X^3 + i X^4,
\end{equation}
identification on the orbits of~\eqref{melisom} is equivalent to the
identification
\begin{equation} \label{melid}
\{X^1, Z, \bar{Z}, \Theta^\Lambda\} 
\sim \{X^1+2\pi R, e^{2\pi i q} Z, 
       e^{-2\pi i q} \bar{Z}, e^{\pi q \Gamma^{\ul3\ul4}} \Theta^\Lambda\}.
\end{equation}
The action is the same as equations~\eqref{Sbos} and~\eqref{SF12}, but
following the identification~\eqref{melid},
the boundary conditions are replaced by
\begin{align}
Z(\sigma+2\pi,\tau) &= e^{2\pi i q w} Z(\sigma, \tau), &
\Theta^\Lambda(\sigma+2\pi,\tau) 
   &= e^{\pi q w \Gamma^{\ul3\ul4}} \Theta^\Lambda(\sigma, \tau),
\end{align}
where $w$ is the winding number in the mode expansion~\eqref{bosmod}.
The effect of the twist is therefore to shift the moding
of $Z$ and
$\Theta$, when
$w\neq0$.  (If $q$ is rational, the moding of $Z$ is not shifted whenever 
$q w \in \ZZ$ and $\Theta$ modes are not shifted when $q w \in 2\ZZ$.)
With the notation
\begin{equation}
\fr{x} \equiv x-\lfloor x \rfloor = \text{the fractional part of $x$},
\end{equation}
then for $\fr{qw}\neq 0$, the mode expansion~\eqref{modXI}
is replaced by
\begin{subequations}
\begin{align}
\begin{split}
Z &= 
\frac{i}{p_- \sqrt{\apr}}
\sum_{n\neq0} \left[
\frac{1}{(n-\fr{q w})\sqrt{c^2+4\frac{\mu^2}{(n-\fr{q w})^2}}} 
\beta_{n-\fr{q w}}
    e^{-i (n-\fr{q w}) 
       \left(\sqrt{c^2+\frac{4\mu^2}{(n-\fr{q w})^2}}\tau+\sigma\right)}
\right. \\ & \qquad \left. 
+ \frac{1}{(n+\fr{q w})\sqrt{c^2+4\frac{\mu^2}{(n+\fr{q w})^2}}} 
\tilde{\beta}_{n+\fr{qw}}
    e^{-i (n+\fr{qw}) 
       \left(\sqrt{c^2+\frac{4\mu^2}{(n+\fr{qw})^2}}\tau-\sigma\right)}
\right],
\end{split} \\ \begin{split}
\bar{Z} &= 
\frac{i}{p_- \sqrt{\apr}}
\sum_{n\neq0} \left[
\frac{1}{(n+\fr{q w})\sqrt{c^2+4\frac{\mu^2}{(n+\fr{q w})^2}}} 
\bar{\beta}_{n+\fr{q w}}
    e^{-i (n+\fr{q w}) 
       \left(\sqrt{c^2+\frac{4\mu^2}{(n+\fr{q w})^2}}\tau+\sigma\right)}
\right. \\ & \qquad \left. 
+ \frac{1}{(n-\fr{q w})\sqrt{c^2+4\frac{\mu^2}{(n-\fr{q w})^2}}} 
\Bar{\tilde{\beta}}_{n-\fr{qw}}
    e^{-i (n-\fr{qw}) 
       \left(\sqrt{c^2+\frac{4\mu^2}{(n-\fr{qw})^2}}\tau-\sigma\right)}
\right],
\end{split}
\end{align}
and if $\fr{\frac{qw}{2}}\neq0$ the expansion~\eqref{fermmodes} is
replaced by
\begin{align}
\begin{split}
& \Theta^1(\sigma,\tau) = 
\sqrt{\frac{c}{2\pi}}
\sum_{n\neq 0} 
e^{\mu \Gamma^{\ul1\ul2} \tau}
\\ & \times
\left[ 
\frac{1}{\sqrt{1+\left(
      \frac{
            \sqrt{(n-i \frfrqw \Gamma^{\ul3\ul4})^2 c^2+4\mu^2}-c}{2\mu}
      \right)^2}}
e^{-i (n-i \frfrqw \Gamma^{\ul3\ul4})
   \left(\sqrt{c^2+\frac{4\mu^2}{(n-i\frfrqw\Gamma^{\ul3\ul4})^2}}\tau-\sigma\right)}
\theta_{n-i\frfrqw \Gamma^{\ul3\ul4}}^1
\right. \\ & \left.
+ \frac{i}{2\mu} 
  \frac{(n+i\frfrqw \Gamma^{\ul3\ul4})
  \left(\sqrt{c^2+\frac{4\mu^2}{(n+i\frfrqw \Gamma^{\ul3\ul4})^2}} - c\right)}{
        \sqrt{1+\left(
            \frac{
            \sqrt{(n+i \frfrqw \Gamma^{\ul3\ul4})^2c^2+4\mu^2}-c}{2\mu}
        \right)^2}}
e^{-i (n+i\frfrqw \Gamma^{\ul3\ul4})
   \left(\sqrt{c^2+\frac{4\mu^2}{(n+i\frfrqw \Gamma^{\ul3\ul4})^2}}\tau
           +\sigma\right)} I 
\theta_{n+i\frfrqw\Gamma^{\ul3\ul4}}^2
\right], 
\end{split} \\ \begin{split}
& \Theta^2(\sigma,\tau) = 
\sqrt{\frac{c}{2\pi}}
\sum_{n\neq 0} 
e^{\mu \Gamma^{\ul1\ul2} \tau}
\\ & \times
\left[ 
\frac{1}{\sqrt{1+\left(
      \frac{
            \sqrt{(n+i \frfrqw \Gamma^{\ul3\ul4})^2 c^2+4\mu^2}-c}{2\mu}
      \right)^2}}
e^{-i (n+i\frfrqw \Gamma^{\ul3\ul4})
   \left(\sqrt{c^2+\frac{4\mu^2}{(n+i\frfrqw \Gamma^{\ul3\ul4})^2}}\tau
           +\sigma\right)}
\theta_{n+i\frfrqw\Gamma^{\ul3\ul4}}^2
\right. \\ & \left.
- \frac{i}{2\mu} 
  \frac{(n-i\frfrqw \Gamma^{\ul3\ul4}) 
     \left(\sqrt{c^2+\frac{4\mu^2}{(n-i\frfrqw \Gamma^{\ul3\ul4})^2}} - c\right)}{
        \sqrt{1+\left(
            \frac{
            \sqrt{(n-i \frfrqw \Gamma^{\ul3\ul4})^2c^2+4\mu^2}-c}{2\mu}
        \right)^2}}
   e^{-i (n-i\frfrqw \Gamma^{\ul3\ul4})
      \left(\sqrt{c^2+\frac{4\mu^2}{(n-i\frfrqw \Gamma^{\ul3\ul4})^2}}\tau
         -\sigma\right)}
I \theta_{n-i\frfrqw\Gamma^{\ul3\ul4}}^1
\right].
\end{split}
\end{align}
\end{subequations}
Although the labels would make more sense in a
(complex) basis of eigenspinors of $\Gamma^{\ul3\ul4}$, the expression is
unambiguous.
The hermiticity properties and commutation relations are essentially
the same as in section~\ref{sec:gsferm12}.

\section{T-duality} \label{sec:Tdual}

Performing a T-duality
of the nine dimensional geometry~\eqref{S12d9}
along the $X^1$-direction leads to the IIA
configuration%
\footnote{The normalizations are that of~\cite{bigbook},
except that, $\rrf_{\text{here}}=\frac{1}{8} \rrf_{\text{\cite{bigbook}}}$.}
\begin{subequations} \label{Tdual}
\begin{align}
ds^2_{\text{IIA}} &= 2 dX^+ dX^- 
- 4 \mu^2 \left[4(X^2)^2+(X^I)^2\right] (dX^+)^2
   + (dX^i)^2, \\
B &= -4 \mu X^2 dX^1 dX^+, \\
{^{(3)}}A &= 8 \mu X^+ dX^2 dX^3 dX^4,
\end{align}
\end{subequations}
where $B$ is the \NSNS\ two-form and ${^{(3)}}A$ is the \RR\ three-form
potential.  It is straightforward to check that these obey the IIA
equations of motion.

This can be further lifted to M-theory, giving the field configuration
\begin{subequations} \label{Mlift}
\begin{align}
ds^2_{\text{M}} &= 2 dX^+ dX^- 
- 4 \mu^2 \left[4(X^2)^2+(X^I)^2\right] (dX^+)^2
   + (dX^i)^2 + (dX^{11})^2, \\
{^{(3)}}C &= 4 \mu X^+ dX^1 dX^2 dX^{11} + 8 \mu X^+ dX^2 dX^3 dX^4.
\end{align}
\end{subequations}
This solution preserves 24 supercharges.
The M-theory Killing spinor equation is~\cite{fp}
\begin{equation} \label{Mkseq}
0 = \covd_\mu \epsilon \equiv \nabla_\mu \epsilon
- \frac{1}{288} {^{(4)}}F_{\sigma\tau\lambda\rho}\left[
    \Gamma^{\sigma\tau\lambda\rho}\Gamma_\mu 
    + 4 \Gamma^{\sigma\tau\lambda} \delta^\rho_\mu\right].
\end{equation}
The integrability condition, $\com{\covd_{\mu}}{\covd_{\nu}} \epsilon=0$
gives $\Gamma^{\ul+} (\Gamma^{\ul1\ul3\ul4\ul{11}}-\one)\epsilon=0$.  Since the
equation~\eqref{Mkseq} is a first-order differential equation, this
means that there are precisely 24 Killing spinors,
namely
\begin{equation} \label{Mkspin}
\epsilon(\psi) = \left[1+\sum_{i=2}^8 X^i \tilde{\Omega}_i\right]
\exp\left[-\frac{\mu}{3} X^+ (3-\Gamma^{\ul-}\Gamma^{\ul+})\Gamma^2
(2\Gamma^{\ul3\ul4}+\Gamma^{\ul1\ul{11}})\right]
 \psi,
\end{equation}
with $\psi$ a constant spinor obeying
$(\Gamma^{\ul1\ul3\ul4\ul{11}}-\one)\Gamma^{\ul+}\psi=0$,
and
\begin{subequations}
\begin{gather}
\tilde{\Omega}_2 = \frac{2}{3}\mu \Gamma^{\ul+} 
      (2\Gamma^{\ul3\ul4}-\Gamma^{\ul1\ul{11}}), 
\\
\begin{align}
\tilde{\Omega}_{I=3,4} &= -\frac{1}{3}\mu \Gamma^{\ul+\ul2\ul I}
      (4 \Gamma^{\ul3\ul4}+\Gamma^{\ul1\ul{11}}) &
\tilde{\Omega}_{I=5,\dots,8} &= \frac{1}{3} \mu \Gamma^{\ul+\ul2\ul I}
               (2\Gamma^{\ul3\ul4}
                -\Gamma^{\ul1\ul{11}}).
\end{align}
\end{gather}
\end{subequations}
The continued presence of the 24 Killing spinors after T-duality may
not seem surprising, but it occurs only because the IIB Killing
spinors do not carry momentum---or equivalently, are independent of
$X^1$, as is shown in appendix~\ref{sec:altkillspin}.  Otherwise,
some of the spinors would have winding and would not be visible in the
T-dualized supergravity, as in~\cite{duff2} (see also the recent~\cite{clp}).
Observe also that these M-theory Killing spinors are similarly
independent of
both $X^1$ and $X^{11}$.

\section{The AdS$_5\times$S$^5$ Origin of the Circle} \label{sec:AdS}

The pp-wave arises from the Penrose limit of
AdS$_5\times$S$^5$~\cite{bfhp2,bfp,bmn}.  This
is a powerful observation, as has been made clear by the CFT
description of~\cite{bmn}.
In~\cite{bfp} it was explained that the Penrose limit preserves the
isometries of the original spacetime (though not the algebra).
Since we have not gained any isometries by taking the Penrose limit,
the isometry on which we are compactifying must correspond to some
isometry of AdS$_5\times$S$^5$; that is, our compactified spacetime
is the Penrose limit of a quotient of AdS$_5\times$S$^5$.  That
quotient will be identified here.

AdS$_5\times$S$^5$ has isometry group SO(4,2)$\times$SO(6), with
generators $M_{\mu\nu}$ and $P_\mu$, where $M_{\mu\nu}$ are rotations
and $P_\mu$ are ``translations'' which commute to rotations.
(For unity of the presentation,
either both $\mu,\nu=0\dots4$ or both $\mu,\nu=5\dots9$.)
In terms of embedding coordinates $Y^M$, where $\eta_{MN}Y^M Y^N = \pm 1$,
the isometries are rotational: with
\hbox{$M_{MN} = Y^M\p_N - Y^N\p_M$} for $M,N=-1\dots4$ or $M,N=5\dots10$,
$P_\mu$ is then $M_{-1,\mu}$ or $M_{10,\mu}$.
In~\cite{bfp,hks}, it was shown that under the Penrose limit,
\begin{align}
P_i &\rightarrow e_{i}, & 
\begin{cases}
M_{0i}, & i=1\dots4 \\
M_{9i}, & i=5\dots8 \end{cases} &\rightarrow 2\mu e_i^*.
\end{align}
Therefore, our quotient by the $\ki{S^+_{12}}$ isometry corresponds to
a quotient by the linear combination $P_1+M_{02}$.  Alternatively, we
could map it to a discrete quotient of the sphere by $P_5+M_{96}$.
Interestingly, Behrndt and L\"{u}st have shown that 
$($AdS$_5/\ZZ_N)\times$S$^5$ orbifolds are U-dual to
AdS$_5\times($S$^5/\ZZ_N)$ orbifolds.\cite{bl}
We see that this obviously holds for $\ZZ$-orbifolds as well, at least
after taking the Penrose limit.  It would be interesting to understand
if this is true before taking the Penrose limit, as well.

On the AdS side, 
this $\ZZ$-quotient is one of those discussed by Ghosh and Mukhi~\cite{gm}.
Specifically, introducing ``light-cone'' coordinates in the
embedding space,
\begin{align}
z_1^\pm &= Y^0 \pm Y^2, &
z_2^\pm &= Y^1 \pm Y^{-1}, &
w &= Y^3 \pm i Y^4,
\end{align}
and the corresponding AdS$_5$ coordinates
\begin{align}
z_1^\pm &= \cosh \frac{\theta_1}{2} e^{\pm \delta}, &
z_2^\pm &= \sinh \frac{\theta_1}{2} \cos \frac{\theta_2}{2} e^{\pm \alpha}, &
w &= \sinh \frac{\theta_1}{2} \sin\frac{\theta_2}{2},
\end{align}
then the quotient is by $\frac{\p}{\p \delta}+\frac{\p}{\p \alpha}$.
This group action is free, and preserves half the supersymmetries.
In section~\ref{sec:compact}, it was shown that after
taking the Penrose limit, it preserves $3/4$ of the supersymmetries.  It
is already known that enhancement of supersymmetry often occurs
upon taking the Penrose limit~\cite{ikm,go,as,kprs,tt,fk}.  Here we
have the novel phenomenon that the supersymmetry is enhanced not to
the full 32 supercharges but to 24.

The more complicated $\ki{S^+_{15}}$ mixes the AdS with the sphere via
$P_1+M_{95}$.  The Melvin twist adds some $M_{34}$ (or $M_{78}$).

\section{Conclusions} \label{sec:conc}

The most general set of spacelike isometries on
which one can compactify the pp-wave has been identified.  In contrast
to Minkowski space, there are a large number of choices for
compactification on supersymmetric circles and tori, even before considering
the shape of the torus.  Some of these are given in
table~\ref{tab:sym}.  Even more remarkably, one finds
compactifications to nine dimensions which preserve 20 or 24
supercharges, and compactifications to eight dimensions which preserve
18 or 20 supercharges.  
Whether these peculiar amounts of supersymmetry arise in a way related
to the
$3/4$ BPS states of~\cite{gh} bears investigating.
T-duality gives IIA and 11-dimensional supergravity solutions that
also preserve precisely 24 supercharges.

It should be noted that
it is easy to see
a very
similar story for the maximally supersymmetric M-theory wave~\cite{fp}.
In particular, using the notation of~\cite{fp}, 
compactification on circles generated
by
$\ki{e_1} + \frac{3}{\mu} \ki{e_2^*}$ and $\ki{e_4} + \frac{6}{\mu}
\ki{e_5^*}$ will preserve 24 supercharges.

Understanding the CFT dual to the compactifications described here
would be very interesting.  In
section~\ref{sec:AdS}, it was seen that the pp-wave compactification arises
from a free, and fairly simple, $\ZZ$-orbifold of AdS$_5$, of a type
discussed in~\cite{gm}.  However, the simpleness of the orbifold could
be quite deceptive; it has been noted in~\cite{hm,gao} that an orbifold
of AdS by a discrete group is typically dual to a CFT
on a non-Hausdorff space.
For this reason, it may make more sense to try to consider, using the
techniques of~\cite{ks}, the
equivalent description of the pp-wave compactification via an
S$^5/\ZZ$.  (Finite orbifolds have recently been discussed in this context
in~\cite{as,kprs,tt,fk}.)

This problem is unlikely to improve for the orbifolding that includes
the Melvin twist.  However the rewards of such an investigation
could include a definitive resolution of the fate of the Melvin tachyon.
(An interesting analysis of a stable twisted compactification of
AdS$_7\times$S$^4$ was given in~\cite{chc}.)

Finally we saw that simultaneous compactification on $S^+_{12}$ and
$S^-_{21}$ results in a noncompact, nonabelian orbifold that preserves
24 supercharges, and includes a null identification.  It would be
interesting to study this orbifold.

\acknowledgments

I thank B.~Acharya, D.~Berenstein,
C.~Hofman, S.~Kachru, H.~Liu, N.~Lambert, J.~Majumder and G.~Moore
for stimulating conversations.
This research was supported in
part by DOE grant \hbox{\#DE-FG02-96ER40559}.

\appendix

\section{More Manifest Isometries} \label{sec:spm}

In the text compactification was considered only along
$\ki{S^+_{ij}}$, but it was noted that it is possible to consider the
$T^2$ compactification along $\ki{S^\pm_{12}}$, say.  This is made
manifest via the coordinate transformation
\begin{subequations}
\begin{gather}
\begin{align}
x^+ &= y^+, &
x^- &= y^- + 2 \mu y^1 y^2 \sin(4\mu y^+), &
x^I &= y^I, 
\end{align} \\ \begin{align}
x^1 &= -(y^1+y^2) \cos(2\mu y^+), &
x^2 &= -(y^1-y^2) \sin(2\mu y^+).
\end{align}
\end{gather}
\end{subequations}
The field configuration in these coordinates is
\begin{subequations}
\begin{align}
ds^2 &= 2 dy^+ dy^- - 4 \mu^2 y^I y^I (dy^+)^2 
+ (dy^1)^2 + (dy^2)^2 + 2 \cos 4 \mu y^+ dy^1 dy^2
+ dy^I dy^I, \\
\rrf &= \frac{1}{8} d\cos(4\mu y^+) dy^1 dy^2 dy^3 dy^4
        + \mu dy^+ dy^5 dy^6 dy^7 dy^8.
\end{align}
\end{subequations}
Note that the metric---and the coordinate transformation---is
singular at $y^+=\frac{n \pi}{4\mu}$.  Of course, these are just
coordinate singularities.  The analysis of section~\ref{sec:compact}
shows that this compactification preserves 16 supercharges.

\section{Killing Spinors in the $X^\mu$ Coordinate System} 
\label{sec:altkillspin}

In the coordinate system~\eqref{myX}, 
and using the zenbein~\eqref{myzen}, the Killing spinors are
\begin{multline}
\epsilon(\psi) = \left[\one+\mu X^1 \Gamma^{\ul+\ul2} 
- \mu X^2 \Gamma^{\ul+\ul1} - i X^i \Omega_i\right]
\left[\cos (\mu X^+) \one + \sin (\mu X^+) \Gamma^{\ul1\ul2}\right]
\\ \times
\left[\cos (\mu X^+) \one -i \sin (\mu X^+) I \right]
\left[\cos (\mu X^+) \one -i \sin (\mu X^+) J\right]
\psi.
\end{multline}
Note that $\Omega_\mu$ has the same form~\eqref{defOmega} in this
coordinate and frame.
It was already noted in section~\ref{sec:compact},
via equation~\eqref{liespin}, that the Killing
vector $\ki{S^+_{12}} = -\frac{\p}{\p X^1}$ preserves 
24 supersymmetries.
In this coordinate system the Lie derivative~\eqref{liespin} with
respect to $\ki{S^+_{12}}$ is just
\begin{equation} \label{nox1inspin}
\lie{\ki{S^+_{12}}} \epsilon(\psi) = -\frac{\p}{\p X^1} \epsilon(\psi)
=\epsilon\left(i\mu\Gamma^{\ul2}(\Gamma^{\ul3\ul4}-i\one)\Gamma^{\ul+}
               \psi\right),
\end{equation}
which (again) vanishes precisely for
$
(\Gamma^{\ul3\ul4}-i\one)\Gamma^{\ul+}\psi=0,
$ 
so the 24 Killing spinors are independent of the $X^1$
coordinate.
\iftoomuchdetail
The last equality of equation~\eqref{nox1inspin} is a little tricky
because it looks like two of the
square brackets should get sign flips.  However, that is
only if the expression is nonzero, at which point (roughly)
$\Gamma^{34}=-i\one$ and the two square brackets simply get
interchanged.
\fi%

Reduced to nine dimensions,
the Killing
spinor equation~\eqref{bfhp:killeq} is
\begin{equation}
\covd_\mu \epsilon =
\left[\nabla_\mu 
+ \frac{i}{4} {^{(2)}}F_{\mu\nu} \Gamma^\nu
+ \frac{1}{6} {^{(4)}}F_{\mu\alpha\beta\gamma} \Gamma^{\alpha\beta\gamma}
- \frac{1}{24}{^{(4)}}F_{\alpha\beta\gamma\delta}
     \Gamma^{\alpha\beta\gamma\delta}{_\mu}
\right] \epsilon = 0.
\end{equation}
\iftoomuchdetail
The last term is a rewriting of $\frac{i}{24}(\ast
F)_{\mu\alpha\beta\gamma\delta} \Gamma^{\alpha\beta\gamma\delta}$ using
the fact that the product of all the nine dimensional
$\Gamma$-matrices gives $i \one$.
\fi%
The integrability condition $\com{\covd_+}{\covd_2}\epsilon=0$ implies
$(\Gamma^{\ul3\ul4}-i\one)\Gamma^{\ul+}\epsilon=0$ which is another
way of demonstrating that there are
twenty-four
supercharges in nine dimensions.  The other components of
the integrability condition vanish.

\end{document}